\documentclass[twocolumn,amssymb,nobibnotes,showpacs,groupaddress,aps,prd]{revtex4}
\usepackage{epsfig,amsmath,graphicx,amssymb}

\def\be{\begin{equation}}
\def\ee{\end{equation}}
\def\bea{\begin{eqnarray}}
\def\eea{\end{eqnarray}}
\def\bes{\begin{subequations}}
\def\ees{\end{subequations}}
\begin{document}
\title{Weak-Light Ultraslow Vector Optical Solitons via Electromagnetically Induced Transparency}
\author{Chao Hang}
\address{Department of Physics, East China Normal University,
Shanghai 200062, China}
\author{Guoxiang Huang}
\address{Department of Physics, East China Normal University,
Shanghai 200062, China}
\date{\today}

\begin{abstract}

We propose a scheme to generate temporal vector optical solitons in
a lifetime broadened five-state atomic medium via
electromagnetically induced transparency. We show that this scheme,
which is fundamentally different from the passive one by using
optical fibers, is capable of achieving distortion-free vector
optical solitons with ultraslow propagating velocity under very weak
drive conditions. We demonstrate both analytically and numerically
that it is easy to realize Manakov temporal vector solitons by
actively manipulating the dispersion and self- and cross-phase
modulation effects of the system.

\end{abstract}


\maketitle


The vector nature of light propagating in a nonlinear medium has led
to the discovery of a novel class of solitons, i. e. vector optical
solitons, which are the solutions of two coupled nonlinear
Schr\"{o}dinger (NLS) equations describing the envelope evolution of
two polarization components of an electromagnetic field. In recent
years, considerable attention has been paid to the
temporal\cite{men,isl,bar,cun,kov,ran} and
spatial\cite{seg,kan,ana,del} vector optical solitons in various
nonlinear systems. Due to their remarkable property, vector optical
solitons have promising applications for the design of new types of
all-optical switches and logic gates\cite{isl1}.

Up to now, most vector optical solitons are produced in passive
media such as optical
fibers\cite{isl,bar,cun,kov,ran,seg,kan,ana,del}, in which far-off
resonance excitation schemes are employed in order to avoid
unmanageable optical attenuation and distortion. However, due to the
lack of distinctive energy levels, the nonlinear effect in such
passive media is very weak, and hence to form vector solitons a very
high input light-power is required. In addition, the lack of
distinctive energy levels and transition selection rules also makes
an active control very difficult. In particular, it is hard to
realize Manakov\cite{manakov} temporal vector optical solitons in
optical fibers because the ratio between self-phase modulation (SPM)
and cross-phase modulation (CPM) is not unity and there is also
detrimental energy exchange between two polarization components due
to the existence of four-wave mixing effect. Manakov vector optical
solitons are of great interest, not only because the coupled NLS
equations describing them have beautiful mathematical properties but
also such solitons may be used to realize all-optical
computing\cite{stei}. Different from spatial Manakov vector optical
solitons, which have been observed more than ten years
ago\cite{kan}, temporal Manakov vector optical solitons have not
been realized in experiment up to now.

In this Letter, we propose a scheme to generate temporal vector
optical solitons in a coherent five-level atomic system via
electromagnetically induced transparency (EIT). This resonant EIT
medium has been recently used to realize polarization qubit phase
gate\cite{ott} and reversible memory devices for photon-polarization
qubit\cite{pet}. We show that two continuous-wave (CW) control laser
fields established prior to the injection of a pulsed probe field
induce a quantum interference effect, which can suppress largely the
absorption of the two orthogonal polarization components of the
probe field.  The scheme suggested here is fundamentally different
from the passive ones due to the existence of distinctive
energy-levels that make an active manipulation on the dispersion and
nonlinear effects of the system possible. In addition, contrary to
all passive schemes the vector optical solitons produced in the
present system may have ultraslow propagating velocity and their
production needs only very weak input power. Furthermore, the
controllability of the present scheme allows us also to realize
easily temporal Manakov vector optical solitons by actively
adjusting the parameters of the system. Notice that scalar ultraslow
optical solitons in EIT media have been investigated
recently\cite{wu,huang,hang}. However, up to now there has been no
study on the ultraslow vector optical solitons in an active optical
medium. Our study represents the first work in this direction and
the results may have potential application in optical information
processing and engineering.

Consider a life-timed broadened five-level system (e. g.  a Zeeman
split atomic gas) interacting with a weak, pulsed, linear-polarized
probe field of central frequency $\omega_p/(2\pi)$ and two strong,
linear-polarized  CW control fields of frequencies
$\omega_{c1}/(2\pi)$ and $\omega_{c2}/(2\pi)$, respectively. The two
polarization components of the probe field drive respectively the
transitions from $|3\rangle\leftrightarrow |2\rangle$ and
$|3\rangle\leftrightarrow|4\rangle$, while the two control fields
drive respectively the transitions from $|1\rangle\leftrightarrow
|2\rangle$ and $|5\rangle\leftrightarrow |4\rangle$ (see Fig. 1(a)).
The atoms are trapped in a cell with the temperature lowed to
$0.5\,\,\mu$K to cancel Doppler broadening and collisions. A
possible arrangement of experimental apparatus is shown in Fig.
1(b).
%
\begin{figure}
\centering
\includegraphics[scale=0.3]{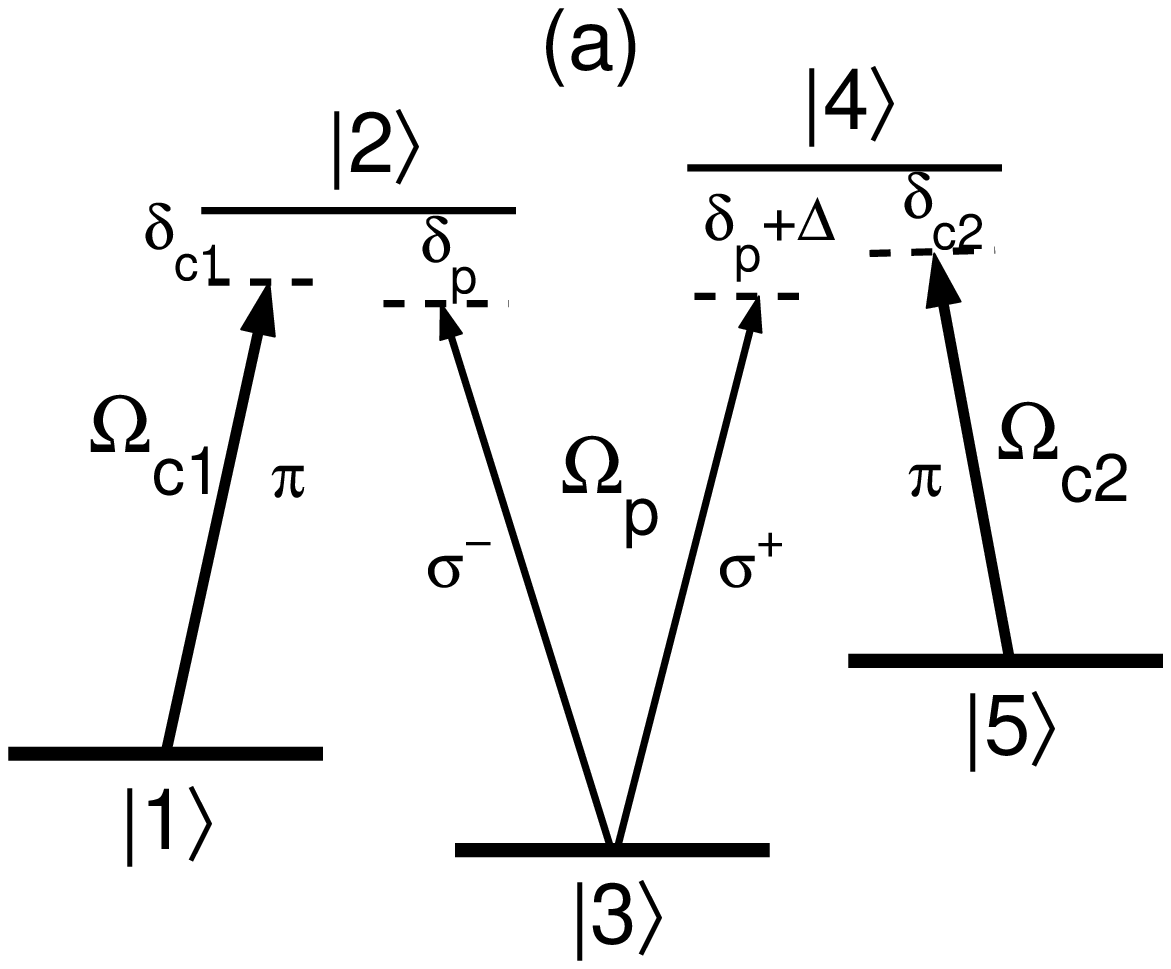}
\includegraphics[scale=0.5]{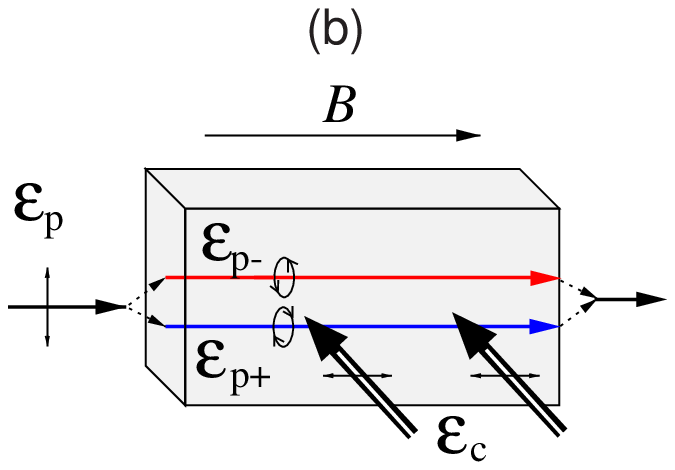}
\caption{\footnotesize{(a): Energy level diagram and excitation
scheme of a five-level atomic system interacting with a weak, pulsed
probe field of Rabi frequency $\Omega_p$ and two strong, CW coupling
fields of Rabi frequency $\Omega_{c1}$ and $\Omega_{c2}$,
respectively. (b): Possible arrangement of experimental apparatus.
}}
\end{figure}
%
The electric-field of the system can be written as ${\bf E}$=$ ({\bf
\hat{\epsilon}}_{+}{\cal E}_{p+} + {\bf \hat{\epsilon}}_{-}{\cal
E}_{p-}) \exp [i (k_p z-\omega_p t)]+{\bf \hat{e}}_{c1} {\cal
E}_{c1} \exp [i (k_{c1} y-\omega_{c1} t)]+ {\bf \hat{e}}_{c2} {\cal
E}_{c2} \exp [i (k_{c2} y-\omega_{c2} t)]+ {\rm c}.{\rm c.}$. Here
${\bf \hat{\epsilon}}_{+}$=$({\bf \hat{x}}+ i{\bf
\hat{y}})/\sqrt{2}$\, (${\bf \hat{\epsilon}}_{-}$=$({\bf \hat{x}}-
i{\bf \hat{y}})/\sqrt{2}$) is the probe-field unit vector of the
$\sigma^{+}$ ($\sigma^{-}$) circular polarization component with the
envelope ${\cal E}_{p+}$ (${\cal E}_{p-})$, which drives the
transition $|2\rangle\leftrightarrow|3\rangle$
($|3\rangle\leftrightarrow|4\rangle$). ${\bf \hat{e}}_{c1}$\, (${\bf
\hat{e}}_{c2}$) is the unit vector of the control field with the
amplitude  ${\bf E}_{c1}$ (${\bf E}_{c2}$), which drives the
transition $|1\rangle\leftrightarrow|2\rangle$
($|4\rangle\leftrightarrow|5\rangle$). Thus the system is composed
of two EIT $\Lambda$-configurations, both of them share the
ground-state level $|3\rangle$\cite{ott,pet}.

In interaction picture, the atomic response of the system under
rotating-wave approximation is described by
\bes \label{AV}
\bea
& & \left(\frac{\partial}{\partial t}+id_1\right)A_{1}
=-i\Omega_{c1}^{\ast}A_2,\label{AV1}\\
& & \left(\frac{\partial}{\partial t}+id_{2n}\right)A_{2n}
=-i\Omega_{cn}A_{4n-3}-i\Omega_{pn}A_3, \label{AV24}\\
& & \left(\frac{\partial}{\partial
t}+id_5\right)A_{5}=-i\Omega_{c2}^{\ast}A_4,\label{AV5}
\eea
\ees
($n$=$1,2$) with $\sum_{j=1}^{5}|A_{j}|^2$=$1$, where $A_j$ is the
probability amplitude of the bare atomic state $|j\rangle$ (with
eigenenergy $\epsilon_j$),
$\Omega_{p1}$=$-(\textbf{p}_{23}$$\cdot$${\bf
\hat{\epsilon}}_{-}){\cal E}_{p-}/\hbar$,
$\Omega_{p2}$=$-(\textbf{p}_{43}$$\cdot$${\bf
\hat{\epsilon}}_{+}){\cal E}_{p+}/\hbar$,
$\Omega_{c1}$=$-(\textbf{p}_{21}$$\cdot$${\bf \hat{e}}_{c1}){\cal
E}_{c1}/\hbar$ and $\Omega_{c2}$=$-(\textbf{p}_{45}$$\cdot$${\bf
\hat{e}}_{c2}){\cal E}_{c2}/\hbar$ are half Rabi frequencies with
$\textbf{p}_{ij}$ being the electric dipole matrix element
associated with the transition from $|j\rangle$ and $|i\rangle$. In
Eq. (\ref{AV}) we have defined
$d_{1}=(\delta_p-\delta_{c1})-i\Gamma_1/2$,
$d_{2}=\delta_p-i\Gamma_2/2$, $d_{3}=-i\Gamma_3/2$,
$d_{4}=(\delta_p+\Delta)-i\Gamma_4/2$ and
$d_{5}=(\delta_p+\Delta-\delta_{c2})-i\Gamma_5/2$ with
$\delta_p=(\epsilon_2-\epsilon_3)/\hbar-\omega_p$,
$\delta_{c1}=(\epsilon_2-\epsilon_1)/\hbar-\omega_{c1}$, and
$\delta_{c2}=(\epsilon_4-\epsilon_5)/\hbar-\omega_{c2}$.
$\Gamma_{j}$ is the decay rate of the state $|j\rangle$,
$\Delta=(2\mu_B/\hbar)gB$ is the Zeeman shift of the upper atomic
sublevel with $\mu_B$ the Bohr magneton, $g$ the gyromagnetic factor
and $B$ the applied magnetic field.

The equation of motion for $\Omega_{pn} (z,t)$ can be obtained by
Maxwell equation under slowly-varying envelope approximation
\be \label{MAX}
i\left(\frac{\partial}{\partial
z}+\frac{1}{c}\frac{\partial}{\partial
t}\right)\Omega_{pn}-\kappa_{3l}
A_{2n}A_{3}^{\ast}=0,\,\,\,\,(l=2n)
\ee
where $\kappa_{32}={\cal N}_a |\textbf{p}_{32}\cdot{\bf
\hat{\epsilon}}_{-}|^2\omega_p/(2\hbar\epsilon_0c)$ and
$\kappa_{34}={\cal N}_a |\textbf{p}_{34}\cdot{\bf
\hat{\epsilon}}_{+}|^2\omega_p/(2\hbar\epsilon_0c)$ with ${\cal
N}_a$ being the atomic density, $\epsilon_{0}$ the vacuum dielectric
constant and $c$ the light speed in vacuum.

Before solving Eqs. (\ref{AV}) and (\ref{MAX}), we examine the
linear properties of the system, which provide main contributors to
pulsed spreading and attenuation. We assume that the probe field is
weak so that the atomic ground state $|3\rangle$ is not depleted,
i.e., $A_{3}\approx$1. Taking $\Omega_{pn}$ and $A_{j}$
($j=1,2,4,5$) as being proportional to $\exp[i(k(\omega)z-\omega
t)]$, one can get two branches of linear dispersion relation
$k_{1}(\omega)$=${\omega}/{c}+\kappa_{32}(\omega-d_{1})/{D_{1}(\omega)}$
and
$k_{2}(\omega)$=${\omega}/{c}+\kappa_{34}(\omega-d_{5})/{D_{2}(\omega)}$,
corresponding to $\sigma^{-}$ and $\sigma^{+}$ components of the
probe field, respectively. Here we have defined
$D_{1}(\omega)$=$|\Omega_{c1}|^{2}-(\omega-d_{1})(\omega-d_{2})$ and
$D_{2}(\omega)$=$|\Omega_{c2}|^{2}-(\omega-d_{4})(\omega-d_{5})$.
%
\begin{figure}
\centering
\includegraphics[scale=0.2]{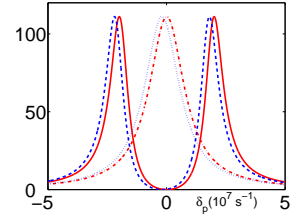}
\caption{\footnotesize{ Absorption spectra of $\Omega_{p1}$ (solid
line) and $\Omega_{p2}$ (dashed line) with parameters
$\Gamma_2\simeq\Gamma_4=2\pi\times6$ MHz, $\Gamma_1\simeq
\Gamma_3\simeq\Gamma_5=2$ kHz,
$\kappa_{32}\simeq\kappa_{34}=1.0\times10^{9}$ cm$^{-1}$s$^{-1}$,
$\Omega_{c1}=\Omega_{c2}=2.0\times10^{7}$ s$^{-1}$, and
$\delta_{c1}=\delta_{c2}=0$, and $\Delta=2.0\times10^{6}$
s$^{-1}$. The dash-dotted (dotted) line represents the absorption
spectra of $\Omega_{p1}$ ($\Omega_{p2}$) when the control fields
are switched off.
 }}
\end{figure}
%
Shown in Fig. 2(a) is the absorption spectra of $\Omega_{p1}$ and
$\Omega_{p2}$.  We see that near the central frequency of the probe
field (i. e. at $\omega$=$0$), both Im($k_{1}$) and Im($k_{2}$) are
close to zero, which means the absorption is greatly suppressed. The
reason of such suppression of the probe field is due to the
introduction of the two strong control fields that induce a quantum
interference effect and thus make the two polarization components of
the probe field transparent.

Because of the existence of dispersion effect, the probe field will
distort during propagation. In the following we show that the SPM
and CPM effects of the system may balance the dispersion. To this
aim we apply the method of multiple-scales\cite{huang,hang} to
investigate the weak nonlinear evolution of the probe field. We make
the asymptotic expansion
$A_{j}$=$\sum^{\infty}_{l=0}\mu^{l}A_{j}^{(l)}$ and
$\Omega_{pn}$=$\sum^{\infty}_{l=1}\mu^{l}\Omega_{pn}^{(l)}$ with
$A^{(0)}_{3}$=$1$ and $A^{(0)}_{j}$=0 ($j$=1,2,4,5), where $\mu$ is
a small parameter characterizing the small population depletion of
the ground state and all quantities on the right hand side of the
asymptotic expansion are considered as functions of the multi-scale
variables $z_{l}$=$\mu^{l}z$ and $t_{l}$=$\mu^{l}t$.

The leading order solution is given by
$\Omega_{pn}^{(1)}=F_n\,\exp \{i[k_n(\omega)z_0-\omega t_0)]\}$
with $F_n$ being yet to be determined envelope functions. At the
second order, a solvability requires $i[
\partial F_n /\partial t_{1}+ V_{gn}\partial F_n /\partial
z_{1} ]=0$, where $V_{gn}$=$1/K_{1n}$ with
$K_{11}=1/c+\kappa_{32}(|\Omega_{c1}|^2+d_1^2)/(|\Omega_{c1}|^2-d_1d_2)^2$
and
$K_{12}=1/c+\kappa_{34}(|\Omega_{c2}|^2+d_5^2)/(|\Omega_{c2}|^2-d_4d_5)^2$.
The solvability condition at the third order yields two coupled
NLS equations
\be \label{ORDER3} i\frac{\partial F_n}{\partial
z_{2}}-\frac{K_{2n}}{2}\frac{\partial^{2} F_n}{\partial
t_{1}^{2}}-(W_{nn}|F_n|^{2}+W_{nm}|F_m|^{2})e^{-2\bar{\alpha}_n
z_2}F_n=0,
\ee
($m,n$=1, 2; $m\neq n$) where
\bea
& & K_{2n}=
-2\kappa_{3l}\frac{d_{2n}|\Omega_{cn}|^2+2d_{4n-3}|\Omega_{cn}|^2+d_{4n-3}^3}
{(|\Omega_{cn}|^2-d_{2n}d_{4n-3})^3},\nonumber\\
 & & W_{nn'}=
-\frac{\kappa_{3l}d_{4n-3}(|d_{4n'-3}|^{2}+|\Omega_{cn'}|^{2})}{D_n|D_{n'}|^{2}},
\nonumber
\eea
($n,n'$=1, 2) are the coefficients characterizing the dispersion
($K_{2n}$), SPM ($W_{nn}$) and CPM ($W_{nm}$, $n\neq m$) of the
two polarization components of the probe field, respectively.
$\bar{\alpha}_n=\mu^2\alpha_n$ with $\alpha=\text{Im}[K_{0n}]$
[$K_{0n}$=$k_n(0)$]. When returning to original variables and
introducing $\delta$=$(1/V_{g1}-1/V_{g2})/2$,
$V_{g}$=$2V_{g1}V_{g2}/(V_{g1}+V_{g2})$, and $\tau=t-z/V_{g}$, Eq.
(\ref{ORDER3}) can be written as the dimensionless form
\bea
& & i\left(\frac{\partial }{\partial
s}+g_{An}\right)u_{n}+(-1)^{n-1}ig_{\delta}\frac{\partial
u_{n}}{\partial \sigma}-\frac{g_{Dn}}{2}\frac{\partial^{2}
u_{n}}{\partial
\sigma^{2}}\nonumber\\
& & -(g_{nn}|u_{n}|^{2}+g_{nm}|u_{m}|^{2})u_{n}=0, \label{vNLS1}
\eea
where $s$=$z/L_D$, $\sigma$=$\tau/\tau_0$, and
$u_{n}$=$(\Omega_{pn}/U_0) e^{-i\tilde{K}_{0n}z}$
($\tilde{K}_{0n}$=Re[$K_{0n}$]), $g_{An}$=$\alpha_nL_D$,
$g_{\delta}$=$\text{sign}(\delta)L_D/L_{\delta}$, $g_{D1}$=
$K_{21}/|K_{22}|$, $g_{D2}$=$\text{sign}(K_{22})$ and
$g_{nn'}$=$W_{nn'}/|W_{22}|$. Here we have defined
$L_D$=$\tau_0^2/|K_{22}|$ (dispersion length), and
$L_{\delta}$=$\tau_0/|\delta|$ (group velocity mismatch length).
$\tau_0$ is typical pulse length of the probe field. In order to
get soliton solutions we have assumed that typical nonlinear
length $L_{NL}=1/(U_0^2|W_{22}|)$ is equal to the dispersion
length $L_D$.

Because coupled NLS Eq. (\ref{vNLS1}) has complex coefficients,
generally a vector soliton does not exist. However, as we show
below that practical parameters can be found based on the EIT
effect and hence the imaginary part of the coefficients can be
much smaller than the corresponding real part. This leads to a
shape-preserving vector optical soliton solution that can
propagate for an extended distance without significant
deformation. The system admits bright-bright, bright-dark, and
dark-dark vector soliton solutions through a balance between the
dispersion and nonlinear effects. The bright-bright vector soliton
solution reads $u_{n}={\cal V}_{n}\rm{sech}\sigma\exp[i({\cal P}_n
\sigma+{\cal Q}_n s)]$ ($n=1,2$) if the parameters fulfill the
condition $g_{22}g_{D1}=g_{12}g_{D2}$. Here we have defined ${\cal
P}_{n}=(-1)^{n-1}g_{\delta}/g_{Dn}$, ${\cal
Q}_{n}=-g_{\delta}^2/(2g_{Dn})-g_{Dn}/2$, and ${\cal
V}_2=[(g_{D1}-g_{11}{\cal V}_1^2)/g_{12}]^{1/2}$.  A bright-dark
vector soliton solution is given by  $u_{1}={\cal
V}_{1}\rm{sech}\sigma\exp[i({\cal P}_1\sigma+{\cal Q}_1 s)]$,
$u_{2}={\cal V}_{2}\tanh\sigma\exp[i({\cal P}_2\sigma+{\cal
Q}_2s)]$, where ${\cal P}_{n}=(-1)^{n-1}g_{\delta}/g_{Dn}$, ${\cal
Q}_{1}=-{\cal P}_1g_{\delta}-g_{D1}(1-{\cal P}_1^2)/2-g_{12}{\cal
V}_2^2$, ${\cal Q}_{2}={\cal P}_2g_{\delta}+g_{D2}{\cal
P}_2^2/2-g_{22}{\cal V}_2^2$, and ${\cal V}_2=[(g_{11}{\cal
V}_1^2-g_{D1})/g_{12}]^{1/2}$.  Here ${\cal V}_1$ is a free
parameter.

Now we show that a realistic atomic system can be found that allows
the bright-bright vector optical soliton described above. We
consider a cold alkali atomic vapor (e. g., rubidium or cesium
atoms) with the decay rates $\Gamma_2\simeq\Gamma_4=0.5\times10^7$
s$^{-1}$, and $\Gamma_1\simeq \Gamma_3\simeq\Gamma_5=1.0\times10^4$
s$^{-1}$. We take $\kappa_{32}\simeq\kappa_{34}=1.0\times10^{9}$
cm$^{-1}$s$^{-1}$ (${\cal N}_a \sim10^{10}$ cm$^{-3}$),
$\Omega_{c1}=\Omega_{c2}=1.6\times10^{8}$ s$^{-1}$,
$\delta_p=1.0\times10^{8}$ s$^{-1}$, $\Delta=2.0\times10^{6}$
s$^{-1}$, $\delta_{c1}=0$, and $\delta_{c2}=3.0\times10^{6}$
s$^{-1}$. With the above parameters, we obtain $K_{01}=-6.41+0.10i$
cm$^{-1}$, $K_{02}=-6.38+0.10i$ cm$^{-1}$,
$K_{11}=(14.62-0.47i)\times10^{-8}$ cm$^{-1}$s,
$K_{12}=(14.72-0.47i)\times10^{-8}$ cm$^{-1}$s,
$K_{21}=(-4.56+0.25i)\times10^{-15}$ cm$^{-1}$s$^{2}$,
$K_{22}=(-4.64+0.26i)\times10^{-15}$ cm$^{-1}$s$^{2}$,
$W_{11}=(-9.37+0.15i)\times10^{-16}$ cm$^{-1}$s$^{2}$,
$W_{12}=(-9.44+0.15i)\times10^{-16}$ cm$^{-1}$s$^{2}$,
$W_{21}=(-9.34+0.15i)\times10^{-16}$ cm$^{-1}$s$^{2}$, and
$W_{22}=(-9.40+0.15i)\times10^{-16}$ cm$^{-1}$s$^{2}$. Notice that
the imaginary parts of these quantities are much smaller than their
relevant real parts. The physical reason for such small imaginary
parts is due to quantum destructive interference induced by two CW
control fields (i. e. EIT effect). We obtain $L_{\delta}=116.8$ cm
and $L_{D}=0.8$ cm with $\tau_0=6.0\times10^{-8}$ s and
$U_0=3.7\times10^7$ s$^{-1}$. The dimensionless coefficients read
$g_{\delta}=0.007$, $g_{D1}=-0.98$, $g_{D2}=-1.0$, and $g_{11}\simeq
g_{12}\simeq g_{21}\simeq g_{22}=-1.0$. The group velocities of the
two polarization components  are respectively given by
Re$(V_{g1})=2.28\times10^{-4}$ $c$ and
Re$({V}_{g2})=2.26\times10^{-4}$ $c$, which means that {\it the two
polarization components of the vector optical soliton propagate with
nearly matched, ultraslow propagating velocities}.

As we have stressed, different from the passive media such as
optical fibers\cite{isl,bar,cun,kov,ran,seg,kan,ana,del} {\it the
parameters of our present EIT medium can be actively manipulated}.
Consequently the coefficients of Eq. (\ref{vNLS1}) can be easily
adjusted to allow us to realize a Manakov system, which is a
completely integrable and can be solved by inverse-scattering
transform\cite{manakov}. In fact, with the parameters given above
Eq. (\ref{vNLS1}) can be written as a near Makakov system ($m,n$=1,
2; $m\neq n$)
\be \label{vNLS2}
i\frac{\partial u_{n}}{\partial s}+\frac{1}{2}\frac{\partial^{2}
u_{n}}{\partial
\sigma^{2}}+(|u_{n}|^{2}+|u_{m}|^{2})u_{n}=R_n(u_n),
\ee
with $R_n(u_n)\simeq-0.08iu_n$ describing the linear absorption
effect. We see that $R_n$ is indeed a small quantity which can be
taken as perturbation. The vector soliton solution of Eq.
({\ref{vNLS2}) after neglecting $R_n$ is
$u_1=\cos\theta\,\rm{sech}(\sigma)\exp(is/2)$ and
$u_2=\sin\theta\,\rm{sech}(\sigma)\exp(is/2)$ with $\theta$ being a
free parameter. Note that since the injected probe field is $\pi$
(i. e. linear) polarized, the two polarization components should
have equal amplitude, i. e. $\theta=\pi/4$.

Shown in Fig. 3 is the evolution  of the two polarization
components of the probe field versus dimensionless time $t/\tau_0$
and distance $z/L_D$. The plots are obtained by numerically
integrating Eq. (\ref{vNLS1}) by using a split-step fast Fourier
transform method and the bright-bright soliton solution given
above as an initial condition. To demonstrate the balance between
the dispersion and nonlinear effects, we change the probe field
amplitude $U_0$ while keep other parameters the same as those
given above. Fig. 3(a) shows the case when the dispersion is
dominant over nonlinearity (i. e.
$U_0\tau_0<\sqrt{|K_{22}/W_{22}|}$). We see that in this case both
polarization components spread and attenuate seriously. Fig. 3(b)
shows that case when $U_0\tau_0$=$\sqrt{|K_{22}/W_{22}|}$, i. e.
there is a balance between the dispersion and nonlinearity. In
this situation a shape-preserving propagation of vector optical
soliton over long distance (z=3.2 cm) is achieved.

%
\begin{figure}
\centering
\includegraphics[scale=0.2]{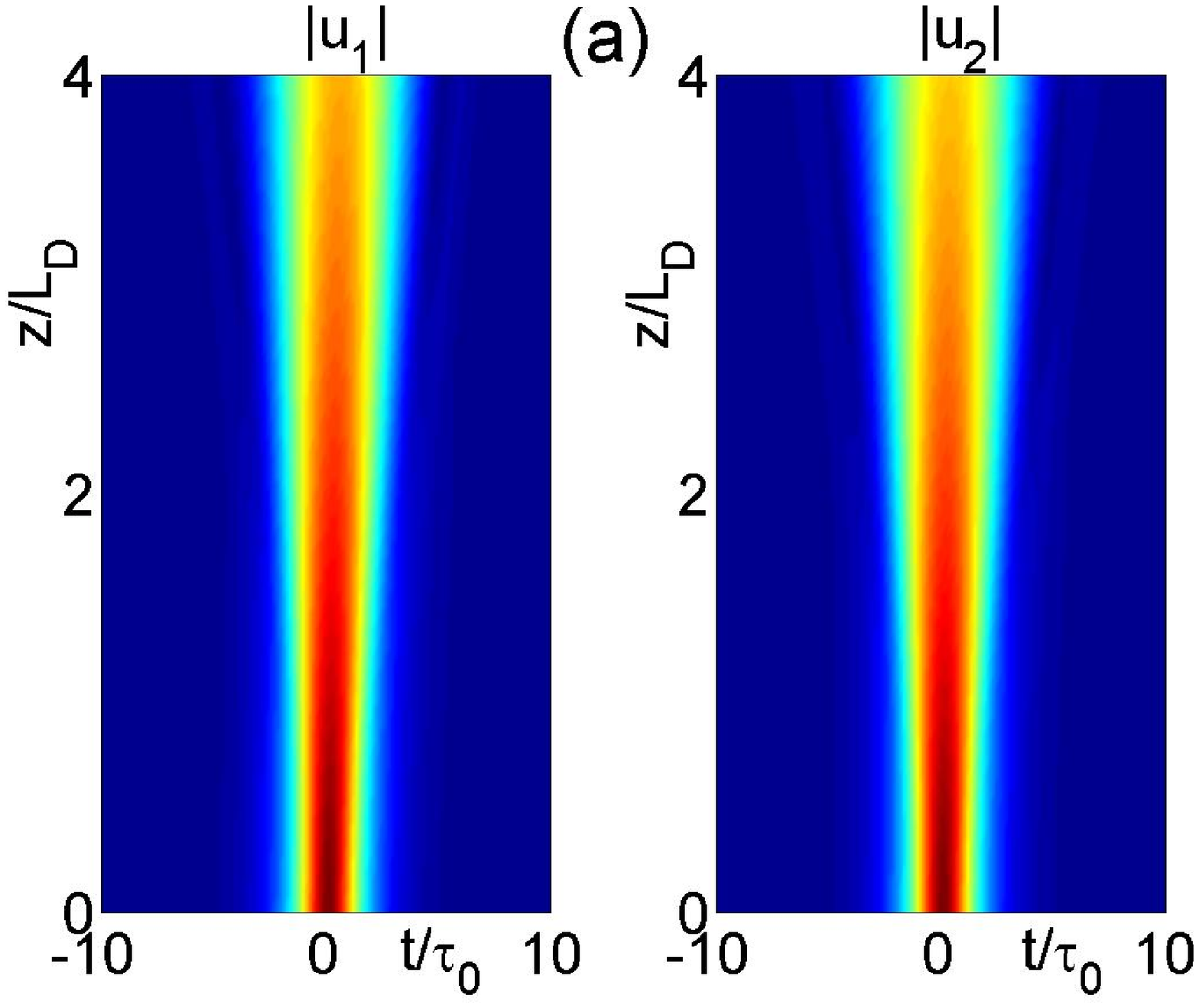}
\includegraphics[scale=0.2]{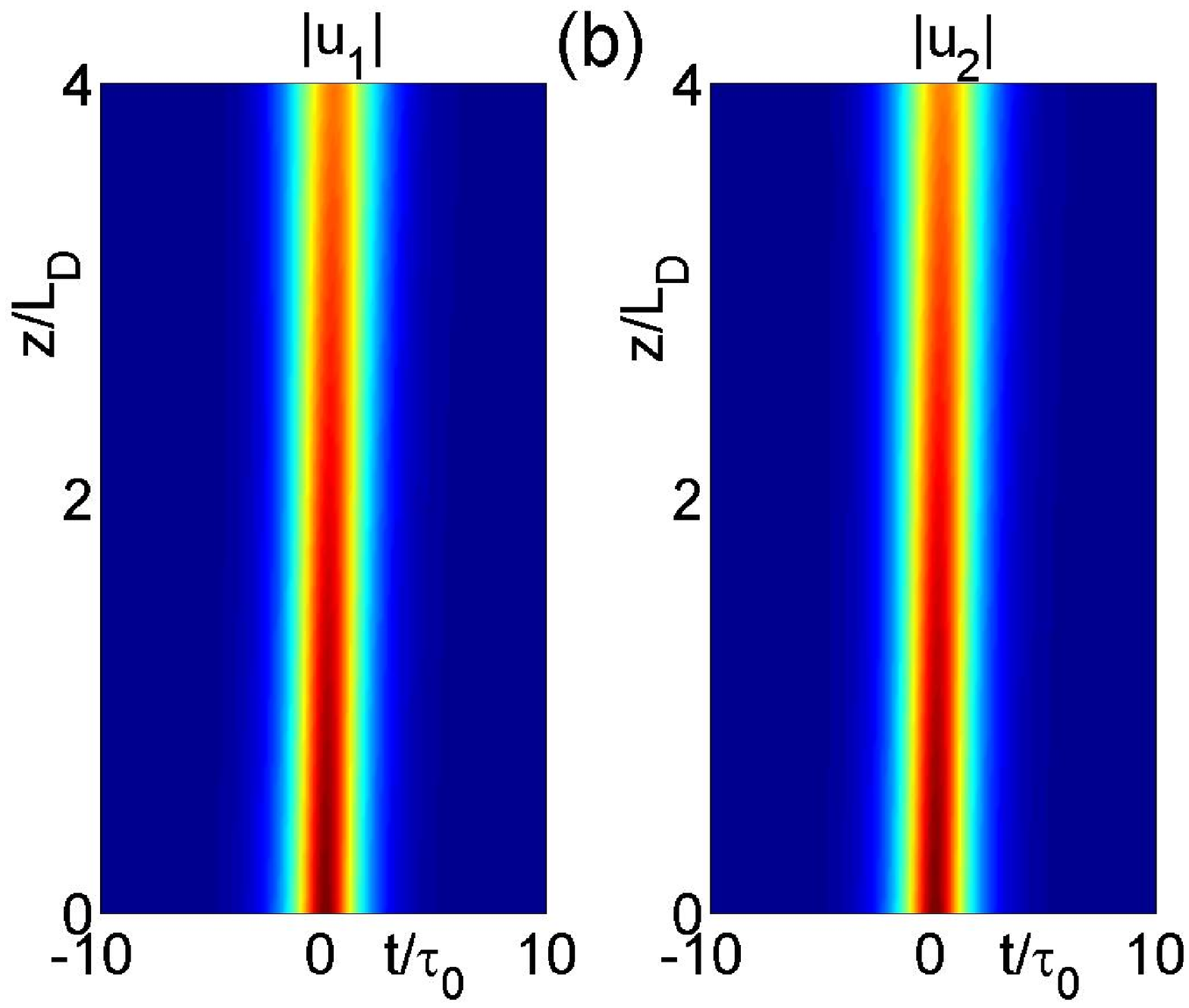}
\caption{ \footnotesize{ Evolution plots of the two polarization
components of the probe field. (a): Dispersion dominant case with
$U_0$=$3.7\times 10^6$ s$^{-1}$. (b): The case of a balance
between dispersion and nonlinearity with $U_0$=$3.7\times 10^7$
s$^{-1}$. Brighter shading marks higher Rabi frequency. The
propagation distance is $z=4L_D=3.2$ cm. } }
\end{figure}
%

The input power of the vector optical soliton can be calculated by
Poynting's vector. It is easy to get the average flux of energy over
carrier-wave period $\bar{P_n}$=$\bar{P_n}_{\rm{max}}\,\text{sech}^2
[ (t-z/V_{gn})/\tau_0]$, with the peak power
$\bar{P_1}_{\rm{max}}\approx\bar{P_2}_{\rm{max}}=8.9\times10^{-4}$
mW. Here we have taken
$|\textbf{p}_{23}|\approx|\textbf{p}_{43}|=2.1\times 10^{-27}$ cm C
and the beam radius of the probe laser  $R_{\perp}=0.01$ cm.  We see
that {\it to generate an ultraslow vector optical soliton in this
active system only very low input power is needed}. This is
drastically different from the vector optical soliton generation
schemes in passive midia\cite{isl,bar,cun,kov,ran,seg,kan,ana,del}
where much higher input power is needed in order to bring out the
nonlinear effect required for the soliton formation.

To make a further confirmation on the vector soliton solutions
obtained and check their stability, we have made additional
numerical simulation directly from Eq. (\ref{AV}) and (\ref{MAX})
without using any approximations. Fig. 4 shows the simulation
result by taking $u_n\,\,(z=0)=2\rm{sech}(2\sqrt{2}\sigma)$
($n=1$, $2$) as initial condition.  We see that the soliton
radiates a small part of energy in its tail but is fairly stable
during propagation.

%
\begin{figure}
\centering
\includegraphics[scale=0.25]{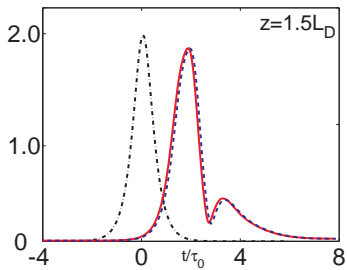}
\caption{\footnotesize{ The bright-bright vector soliton evolution
obtained by integrating directly from Eq. (\ref{AV}) and
(\ref{MAX}) without any approximations.  The solid (dashed) line
is for the $\sigma_-$ ($\sigma_+$) component of the relative probe
intensity $|\Omega_{p1}/U_0|^2$ ($|\Omega_{p2}/U_0|^2$). The
propagation distance is $z=1.5L_D=1.2$ cm. }}
\end{figure}
%

In conclusion, we have proposed a scheme to create temporal vector
optical solitons in a coherent five-level atomic system. Such
solitons can have ultraslow propagating velocity and may be produced
with extremely low input power. We have demonstrated both
analytically and numerically that it is easy to realize Manakov
temporal vector optical solitons by actively manipulating the
dispersion and nonlinear effects of the system. Due to the robust
propagation nature, the ultraslow vector optical solitons suggested
here may have potential application in optical information
processing and engineering under a weak-light level.


This work was supported by the NSF-China under Grant Nos. 90403008
and 10674060, and by the PhD Program Scholarship Fund of ECNU 2006.



\end{document}